\documentstyle[spie,psfig]{article}
\newcommand{\bmt}{\begin{mathletters}}
\newcommand{\emt}{\end{mathletters}}
\newcommand{\lb}[1]{\label{eq:#1}}
\newcommand{\eq}[1]{Eq.~(\ref{eq:#1})}
\newcommand{\ef}[1]{(\ref{eq:#1})}

\newcommand{\nn}{\nonumber}

\newcommand{\wc}{\frac{\w^2}{c^2}}

\newcommand{\bqe}{\begin{eqnarray}}
\newcommand{\eqe}{\end{eqnarray}}

\newcommand{\w}{\omega }
\newcommand{\e}{\epsilon}

\newcommand{\la}{\langle}
\newcommand{\ra}{\rangle}

\newcounter{tr}
\newcommand{\sctr}[1]{\setcounter{tr}{#1}}
\newcommand{\sceq}{\addtocounter{equation}{-1}}

\authorinfo{Other author information:
(Send correspondence to A.A.M)\\A.A.M.:
Email: aamaradu@uci.edu; Telephone: 949-824-5943; Fax: 949-824-2174}
\title{The Angular Intensity Correlation Functions
C$^{(1)}$ and C$^{(10)}$ for the Scattering of S-Polarized Light from a
One-Dimensional Randomly Rough Dielectric Surface}
\author{Ingve Simonsen\supit{a,b}, Alexei A. Maradudin\supit{b}, and Tamara A.
Leskova\supit{c}
\skiplinehalf
\supit{a}Department of Physics,  Theoretical Physics Group,\\
The Norwegian University of Science and Technology,\\ 
N-7491 Trondheim, Norway
\skiplinehalf
\supit{b}Department of Physics and Astronomy\\ and Institute for Surface and
Interface Science,\\
University of California, Irvine, CA \hspace{0.5em}92697 %
\hspace{0.5em} USA \skiplinehalf
\supit{c}Institute of Spectroscopy, Russian Academy of Sciences,\skipline
Troitsk %
\hspace{0.5em}142092\hspace{0.5em} Russia}
\begin{document}
\maketitle

\begin{abstract}
We calculate the short--range contributions $C^{(1)}$ and
$C^{(10)} $ to the angular intensity correlation
function
for the scattering of $s$-polarized light
from a one-dimensional random interface between two dielectric
media.
The calculations are carried out on the basis of a new
approach that separates out explicitly the contributions $C^{(1)}$ and $%
C^{(10)}$ to the angular intensity correlation
function. The contribution $C^{(1)}$ displays peaks associated with the
memory effect and the
reciprocal memory effect.
In the case of a dielectric--dielectric interface,
which does not support
surface electromagnetic surface waves, these peaks arise from the coherent
interference of multiply-scattered lateral waves supported by the interface.
The contribution $C^{(10)}$
is a structureless function of its
arguments.
\end{abstract}


\section*{1. Introduction}

Recent theoretical studies [1,2] of angular intensity correlation functions
of light scattered incoherently from a weakly rough random metal surface have
predicted new correlation functions in addition to the short--range ($C^{(1)}$%
), long--range ($C^{(2)}$), and infinite--range ($C^{(3)}$) correlation
functions predicted in earlier studies of angular intensity correlation
functions of light scattered from volume disordered media [3], and
subsequently observed experimentally [4-6]. The new correlation functions
predicted in the context of rough surface scattering have been labeled the $C^{(10)} $ and $C^{(1.5)}$ correlation functions. The former is of the same
order of magnitude as the $C^{(1)}$ correlation function, which describes
the memory effect and the reciprocal memory effect, so named because of the
wave vector constraints that govern their occurrence. The $C^{(10)}$
correlation effect had been overlooked in an earlier study of angular
intensity correlation functions in rough surface scattering [7] because it
was based on the factorization approximation [8], an approximation that was
not used in Ref. [1, 2]. The $C^{(1.5)}$ correlation function, which also
cannot be obtained within the factorization approximation, was predicted to
display peaks associated with the roughness-induced excitation of the
surface plasmon polaritons supported by the metal surface, just as the $%
C^{(1)}$ correlation function does when the conditions for the occurrence
of the memory effect and the reciprocal memory effect are satisfied. However
the $C^{(1.5)}$ correlation function is significantly weaker than the $%
C^{(1)}$ correlation function. In very recent experimental work [9] the
envelopes of the $C^{(1)}$ and $C^{(10)}$ correlation functions have been
measured experimentally for the scattering of p-polarized light from weakly
rough, one-dimensional gold surfaces. The $C^{(1.5)}, C^{(2)}$, and $C^{(3)}$
correlation functions have yet to be observed experimentally.

In contrast with the theoretical and experimental studies of angular
intensity correlation functions of light scattered from randomly rough metal
surfaces cited above, there have been no theoretical studies to date of
these correlation functions for light scattered from dielectric surfaces, and
the only experimental studies [10] of these correlation functions, which are
limited to $C^{(1)}$, demonstrate the existence of the memory effect and of
the reciprocal memory effect, but do not present results for the envelope of
$C^{(1)}$.

In this paper we present a theoretical study of the angular intensity
correlation functions $C^{(1)}$ and $C^{(10)}$ of s-polarized light
scattered from a one-dimensional random interface between two
dielectric media. Although a dielectric-dielectric interface does not support
surface electromagnetic waves, it does support lateral waves [11], and one of
our aims in this work is to show that these waves can give rise to features
in $C^{(1)}$, namely peaks when the conditions for the occurrence of the memory and the
reciprocal memory effects are satisfied, just as surface plasmon polaritons
do in the case of a vacuum-metal interface. A second aim of this work is to
present an approach to the calculation of an angular intensity correlation
function that explicitly separates out the contributions $C^{(1)}$ and $
C^{(10)}$ to it, allowing each to be calculated separately. The remainder of
the correlation function, which represents the sum of the contributions
labeled $C^{(1.5)}, C^{(2)}$, and $C^{(3)}$, also allows each of these
contributions to be treated separately, and will be the subject of a
separate paper.

\setcounter{section}{2}
\section*{2.  The Scattering System}

The physical system we consider in this work consists of a dielectric medium,
characterized by a real positive dielectric constant $\epsilon _0$, in the
region $x_3>\zeta (x_1)$ and a dielectric medium, characterized by a real
positive dielectric constant $\epsilon $,
\vskip-2.in
{\vbox{\psfig{figure=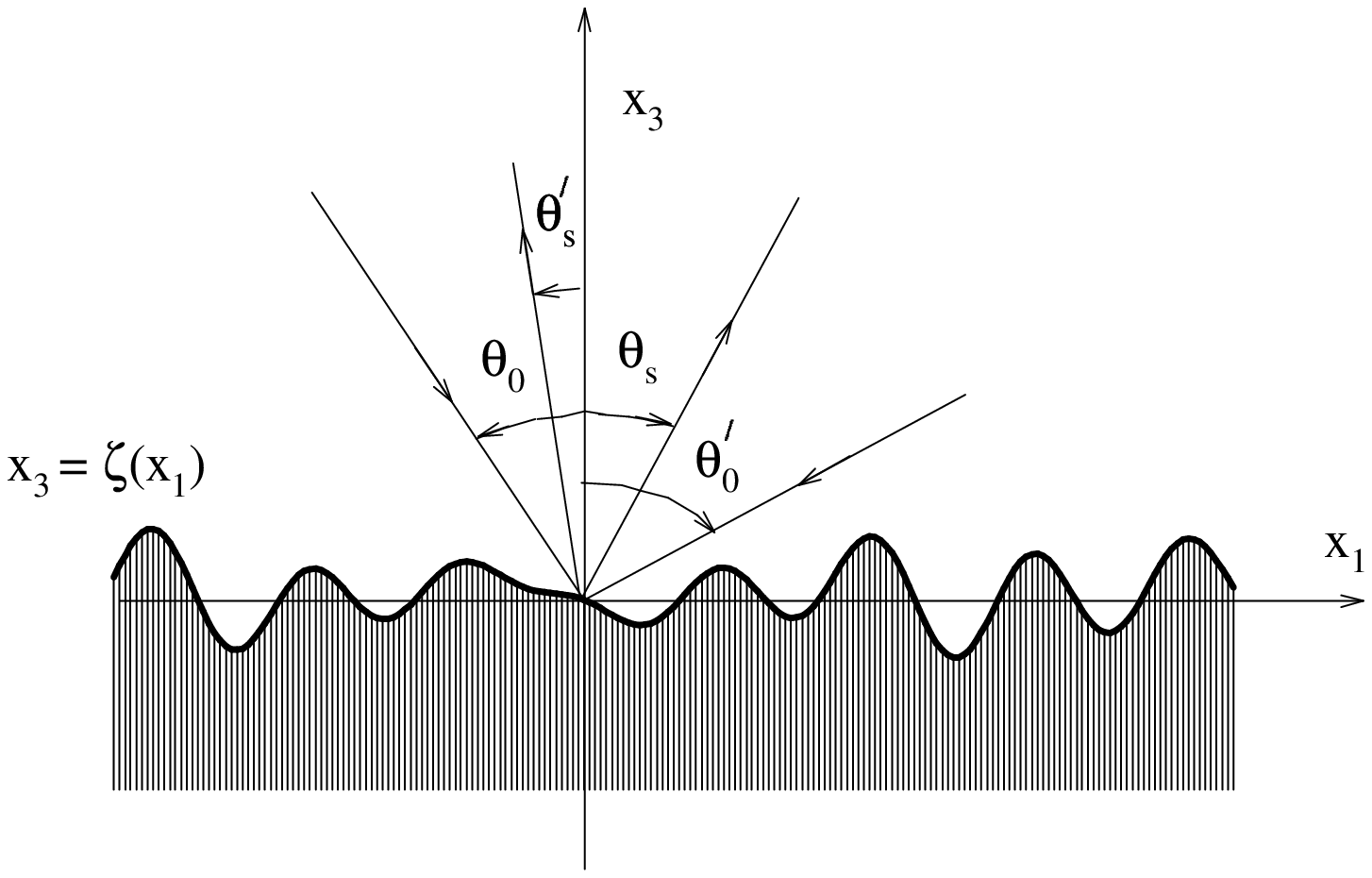,height=4.in,width=7.in}}}
\vskip-2.5in
\begin{itemize}
\item[Fig.1]{The scattering system studied in this paper.}
\end{itemize}
in the region $x_3<\zeta (x_1)$
(Fig. 1).
The surface profile function $\zeta (x_1)$ is assumed to be a
single-valued function of $x_1$ that is differentiable and that constitutes
a zero-mean, stationary, Gaussian random process defined by the properties %
\sctr{0}
\begin{eqnarray}
\langle \zeta (x_1)\rangle  =0, \,\, \,\,\,\,\langle \zeta (x_1)\zeta
(x_1^{\prime })\rangle  &=&\delta ^2W(|x_2-x_1^{\prime }|).
\lb{zeta}
\end{eqnarray}
In Eqs.\ef{zeta} the angle brackets denote an average over the ensemble of
realizations of $\zeta (x_1)$, and $\delta =\langle \zeta ^2(x_1)\rangle ^{%
\frac 12}$ is the rms height of the surface. In numerical calculations we
will use the Gaussian form
\sctr{0}
\begin{eqnarray}
W(|x_1|)=\exp (-x_1^2/a^2)
\lb{w(x)}
\end{eqnarray}
for the surface height autocorrelation function. The characteristic length $a
$ appearing in the expression is called the transverse correlation length of
the surface roughness.

The Fourier coefficient $\hat{\zeta} (k)$ of the surface profile function
is also a zero-mean Gaussian random process characterized by the following
statistical properties: %
\sctr{0}
\begin{eqnarray}
\langle \hat{\zeta} (k) \rangle = 0, \,\,\,\,\,
\langle \hat{\zeta} (k)\hat{
\zeta} (k^{\prime})\rangle = 2\pi\delta (k+k^{\prime})\delta^2g(|k|) ,
\end{eqnarray}
where $g(|k|)$, the power spectrum of the surface roughness, is defined by %
\sctr{0}
\begin{eqnarray}
g(|k|) = \int^{\infty}_{-\infty} dx_1 W(|x_1|)e^{-ikx_1} .
\lb{ps}
\end{eqnarray}
The form of $g(|k|)$ that corresponds to the choice \ef{w(x)}
for $W(|x_1|)$ is
\begin{eqnarray}
g(|k|) = \sqrt{\pi} a\exp (-a^2/k^2/4) .
\lb{gaus}
\end{eqnarray}

\setcounter{section}{3}
\setcounter{equation}{0}
\section*{3.  Scattering Theory}

We assume that the surface $x_3=\zeta (x_1)$ is illuminated from the medium
whose dielectric constant is $\epsilon _0$ by an s-polarized electromagnetic
wave of frequency $\omega $, whose plane of incidence is the $x_1x_3$-plane.
In the  region $x_3>\zeta (x_1)_{max}$ the single nonzero component of the
electric vector of this wave is the sum of an incident wave and a scattered
field,
\begin{eqnarray}
E_2^{>}(x_1,x_3|\omega ) =\exp [ikx_1-i\alpha _0(k)x_3]
+\int_{-\infty }^\infty \frac{dq}{2\pi }R(q|k)\exp [iqx_1+i\alpha
_0(q)x_3],
\lb{inc}
\end{eqnarray}
where
$\alpha _0(q)=\sqrt{\epsilon _0(\omega ^2/c^2)-q^2}$,
$Re \alpha_0(q)>0$, $Im\alpha_0(q)>0$.
A time dependence of the form of $\exp (-i\omega t)$ has been assumed in
 Eq. \ef{inc}  but explicit reference to it has been
suppressed.
The scattering amplitude $R(q|k)$ satisfies the
reduced Rayleigh equation
\begin{eqnarray}
\int_{-\infty }^\infty \frac{dq}{2\pi }M(p|q)R(q|k)=N(p|k),
\lb{rre}
\end{eqnarray}
where
\sctr{0}
\begin{eqnarray}
M(p|q) =\frac{I(\alpha (p)-\alpha _0(q)|p-q)}
{\alpha (p)-\alpha _0(q)}, \, \,\,
N(p|k) =-\frac{I(\alpha
(p)+\alpha _0(k)|p-k)}{\alpha (p)+\alpha _0(k)},
\end{eqnarray}
with $\alpha (q)=\sqrt{\epsilon (\omega ^2/c^2)-q^2}$,
$Re \alpha(q)>0$ , $Im \alpha(q)>0$,  and
\sctr{0}
\begin{eqnarray}
I(\gamma |Q)=\int_{-\infty }^\infty dx_1\,e^{-iQx_1}e^{-i\gamma \zeta (x_1)}.
\end{eqnarray}
The solution of \eq{rre} can be written in the form [12]
\sctr{0}
\begin{eqnarray}
R(q|k)=-2\pi \delta (q-k)-2iG(q|k)\alpha _0(k),
\lb{rqk}
\end{eqnarray}
where $G(q|k)$  is the Green's function  associated with the random  rough
interface. It is defined as the
solution of the equation
\begin{eqnarray}
G(q|k) &=& 2\pi\delta (q-k)G_0(k) +  \int^{\infty}_{-\infty}
\frac{dp}{2\pi} G_0(q)V(q|p)G(p|k),
\lb{gr}
\end{eqnarray}
where
\begin{eqnarray}
G_0(k)=\frac i{\alpha _0(k)+\alpha (k)}
\lb{g0}
\end{eqnarray}
is the Green's function for a planar dielectric-dielectric interface.
The scattering potential $V(q|k)$
is defined by Eqs. \ef{rre} and \ef{gr}, and is found to satisfy
the integral equation
\begin{eqnarray}
\int_{-\infty }^\infty \frac{dp}{2\pi }[M(q|p)+N(q|p)]\frac{V(p|k)}{%
2i\alpha _0(p)}
=\{M(q|k)[\alpha _0(k)-\alpha (k)]-N(q|k)[\alpha
_0(k)+\alpha (k)]\}\frac 1{2\alpha _0(k)}.
\lb{vsc}
\end{eqnarray}

In what follows we will need the averaged Green's function $\langle
G(q|k)\rangle$. Due to the stationarity of the surface profile function $\zeta(x_1)$, $\langle G(q|k)\rangle$ must be diagonal in $q$ and $k$,
$\langle G(q|k)\rangle = 2\pi\delta (q-k)G(k).
$
An application of the smoothing method [12] to  \eq{gr} yields the result
that $G(k)$ is given by
\begin{eqnarray}
G(k) = \frac{1}{G^{-1}_0(k)-M(k)} = \frac{i}{\alpha_0 (k) + \alpha (k) -
i M(k)} .
\lb{dy}
\end{eqnarray}
The proper self-energy $M(k)$ appearing in these expressions is defined by
the relation
$\langle M(q|k) \rangle = 2\pi \delta (q-k)M(k) ,
$ where the unaveraged self-energy $M(q|k)$ is the solution of %
\begin{eqnarray}
M(q|k) = V(q|k) + \int^{\infty}_{-\infty} \frac{dp}{2\pi} M(q|p)
G_0(p)[V(p|k)-\langle V(p|k)\rangle ] .
\lb{self}
\end{eqnarray}

In order to calculate $M(k)$ and the Green's function $G(q|k)$
we
need the scattering potential $V(q|k)$. We can solve the equation it
satisfies, \eq{vsc}, as an expansion in powers of the surface profile
function $\zeta (x_1)$. In what follows we will make the small roughness
approximation [13]. This consists of approximating $V(q|k)$ by the term of
first order in the surface profile function $V^{(1)}(q|k)$:
\sctr{0}
\begin{eqnarray}
V^{(1)}(q|k)=(\epsilon -\epsilon _0)\frac{\omega ^2}{c^2}\hat{\zeta}(q-k).
\lb{v1}
\end{eqnarray}
The results we obtain are therefore limited to weakly rough surfaces. In the
small roughness approximation the self-energy $M(k)$ obtained by  \eq{self}
is given to lowest nonzero order in  $\zeta (x_1)$ by
\begin{eqnarray}
M(k)=W^2\int_{-\infty }^\infty \frac{dp}{2\pi }g(|k-p|)G_0(p),
\lb{m(k)}
\end{eqnarray}
where
\begin{eqnarray}
W=\delta (\epsilon -\epsilon _0)\frac{\omega ^2}{c^2}.
\end{eqnarray}

\setcounter{equation}{0}
\setcounter{section}{4}
\section*{4.  The Correlation Function $C(q,k|q',k)$}

The angular intensity correlation function of interest to us is defined by
\begin{eqnarray}
C(q,k|q^{\prime },k^{\prime })=\langle I(q|k)I(q^{\prime }|k^{\prime
})\rangle -\langle I(q|k)\rangle \langle I(q^{\prime }|k^{\prime })\rangle .
\lb{cin}
\end{eqnarray}
The intensity $I(q|k)$ entering this expression is defined in terms of the
scattering matrix $S(q|k)$ for the scattering of light of frequency $\omega $
from a one-dimensional random surface by
\sctr{1}
\begin{eqnarray}
I(q|k)=\frac {\sqrt{\e_0}}{L_1}\left( \frac \omega c\right) |S(q|k)|^2,
\lb{i(q,k)}
\end{eqnarray}
where $S(q|k)$ is given in terms of the scattering amplitude $R(q|k)$ by %
\sctr{2}\sceq
\begin{eqnarray}
S(q|k)=\frac{\alpha _0^{\frac 12}(q)}{\alpha _0^{\frac 12}(k)}R(q|k).
\lb{s(q,k)}
\end{eqnarray}
The wavenumbers $k$ and $q$ in Eqs. \ef{i(q,k)} and \ef{s(q,k)}
are related to the
angles of incidence and scattering, $\theta _0$ and $\theta _s$,
respectively, by (Fig. 1)
\sctr{0}
\begin{eqnarray}
k=\sqrt{\epsilon _0}(\omega /c)\sin \theta _0,\,\,\,q=\sqrt{\epsilon _0}%
(\omega /c)\sin \theta _s;
\end{eqnarray}
$L_1$ is the length of the $x_1$-axis covered by the random surface; and the
angle brackets in \eq{cin} denote an average over the ensemble of
realizations of the surface profile function $\zeta (x_1)$.

We note that the definition of the intensity used by West and O'Donnell in
their study of angular intensity correlation functions in the
scattering of light from a random vacuum-metal interface [9] differs by a
factor from that given by \eq{i(q,k)}, namely
\begin{eqnarray}
I(q|k)_{\mathrm{wo}}=\frac{\cos \theta _s}{2\pi\sqrt{\e_0} }I(q|k).
\end{eqnarray}

With the definition of $I(q|k)$ given by \eq{i(q,k)} we write the correlation
function $C(q,k|q^{\prime },k^{\prime })$ in the form
\begin{equation}
C(q,k|q^{\prime },k^{\prime })=\e_0\frac 1{L_1^2}\left( \frac \omega c\right) ^2%
\frac{\alpha _0(q)\alpha _0(q^{\prime })}{\alpha _0(k)\alpha _0(k^{\prime })}%
[\langle |R(q|k)|^2|R(q^{\prime }|k^{\prime })|^2\rangle -\langle
|R(q|k)|^2\rangle \langle |R(q^{\prime }|k^{\prime })|^2\rangle ].
\lb{cr}
\end{equation}
If we substitute into this expression the result for $R(q|k)$ given by
\eq{gr}, and omit all terms proportional to $\delta (q-k)$ and/or $\delta
(q^{\prime }-k^{\prime })$, as uninteresting specular contributions, we obtain
\begin{eqnarray}
C(q,k|q^{\prime },k^{\prime }) = \epsilon_0 \frac{16\e_0}{L_1^2}\left( \frac \omega c%
\right) ^2\alpha _0(q)\alpha _0(k)\alpha _0(q^{\prime })\alpha _0(k^{\prime
})
 [\langle |G(q|k)|^2|G(q^{\prime }|k^{\prime })|^2\rangle -\langle
|G(q|k)|^2\rangle \langle |G(q^{\prime }|k^{\prime })|^2\rangle ].
\lb{cg}
\end{eqnarray}

Now $G(q|k)$ satisfies [14]
\begin{eqnarray}
G(q|k)=2\pi \delta(q-k) G(k) +G(q)t(q|k)G(k),
\lb{gt}
\end{eqnarray}
where the function $t(q|k)$ is the solution of [14]
\begin{eqnarray}
t(q|k)=w(q|k)+\int_{-\infty }^\infty \frac{dp}{2\pi }w(q|p)G(p)t(p|k),
\lb{t(q,k)}
\end{eqnarray}
with
\begin{eqnarray}
w(q|k)=V(q|k)-\langle M(q|k)\rangle .
\lb{w(q,k)}
\end{eqnarray}
Then, again omitting all terms proportional to $\delta (q-k)$ and/or $\delta
(q^{\prime }-k^{\prime })$, we obtain from Eq. (4.6)
\begin{eqnarray}
C(q,k|q^{\prime },k^{\prime }) =  \epsilon_0\frac{16\e_0}{L_1^2}\left( \frac \omega c%
\right) ^2\alpha _0(q)|G(q)|^2\alpha _0(k)|G(k)|^2
D(q,k|q^{\prime },k^{\prime })\alpha _0(q^{\prime })|G(q^{\prime
})|^2\alpha _0(k^{\prime })|G(k^{\prime })|^2,
\lb{ct}
\end{eqnarray}
where
\begin{equation}
D(q,k|q^{\prime },k^{\prime })=\langle t(q|k)t^{*}(q|k)t(q^{\prime
}|k^{\prime })t^{*}(q^{\prime }|k^{\prime })\rangle -\langle
t(q|k)t^{*}(q|k)\rangle \langle t(q^{\prime }|k^{\prime })t^{*}(q^{\prime
}|k^{\prime })\rangle .
\lb{d}
\end{equation}

Before  proceeding  we also introduce the normalized angular intensity
correlation function defined by
\begin{eqnarray}
\Xi (q,k|q^{\prime },k^{\prime })=\frac{\langle I(q|k)I(q^{\prime
}|k^{\prime })\rangle -\langle I(q|k)\rangle \langle I(q^{\prime }|k^{\prime
})\rangle }{\langle I(q|k)\rangle \langle I(q^{\prime }|k^{\prime })\rangle }%
,
\lb{norm}
\end{eqnarray}
which in terms of the  function $t(q|k)$ takes the form
\begin{eqnarray}
\Xi (q,k|q^{\prime },k^{\prime })=\frac{D(q,k|q^{\prime },k^{\prime })}{%
\langle |t(q|k)|^2\rangle \langle |t(q^{\prime }|k^{\prime })|^2\rangle }.
\lb{normt}
\end{eqnarray}
We note that this definition of the normalized angular intensity correlation
function differs from that introduced by West and O'Donnell [9].
If we note that $\langle t(q|k)\rangle =0$ (see \eq{gt}), we can rewrite $%
D(q,k|q^{\prime },k^{\prime })$ equivalently as
\begin{equation}
D(q,k|q^{\prime },k^{\prime })=|\langle t(q|k)t^{*}(q^{\prime }|k^{\prime
})\rangle |^2+|\langle t(q|k)t(q^{\prime }|k^{\prime })\rangle
|^2+\{t(q|k)t^{*}(q|k)t(q^{\prime }|k^{\prime })t^{*}(q^{\prime }|k^{\prime
})\},
\lb{dt}
\end{equation}
where $\{\cdots \}$ denotes the cumulant average [15].

The result given by \eq{dt} is very convenient. Due to the stationarity
of the surface profile function $\zeta (x_1)$, $\langle t(q|k)t^{*}(q^{\prime
}|k^{\prime })\rangle $ is proportional to $2\pi \delta (q-k-q^{\prime
}+k^{\prime })$. It therefore gives rise to the contribution to $%
C(q,k|q^{\prime },k^{\prime })$ called $C^{(1)}(q,k|q^{\prime },k^{\prime })$
, and describes the memory effect and the reciprocal memory effect.
Similarly, $\langle t(q|k)t(q^{\prime }|k^{\prime })\rangle $ is
proportional to $2\pi \delta (q-k+q^{\prime }-k^{\prime })$, and therefore
contributes the correlation function $C^{(10)}(q,k|q^{\prime },k^{\prime })$
to $C(q,k|q^{\prime },k^{\prime })$. The third term on the right hand side
of \eq{dt}  $\{ t(q|k)t^*(q|k)t(q'|k')t^{*}(q^{\prime }|k^{\prime })\} $ due
to the stationarity of the surface profile function $\zeta (x_1)$ is
proportional to $2\pi\delta (0)= L_1$, and  gives rise to the long- and infinite-range contributions to $C(q,k|q^{\prime },k^{\prime })$ given by the sum $%
C^{(1.5)}(q,k|q^{\prime },k^{\prime })+C^{(2)}(q,k|q^{\prime },k^{\prime
})+C^{(3)}(q,k|q^{\prime }|k^{\prime })$. Thus, the approach presented here
separates explicitly the contributions to $C(q,k|q^{\prime },k^{\prime })$
that have been named $C^{(1)}(q,k|q^{\prime },k^{\prime })$ and $%
C^{(10)}(q,k|q^{\prime },k^{\prime })$. What is more, this approach
explicitly shows the relative magnitudes of the different contributions to
the correlation function. Indeed,
since $2\pi\delta (0)=L_1,$ when the arguments of the delta-functions vanish
 the $C^{(1)}(q,k|q^{\prime },k^{\prime })$ and $%
C^{(10)}(q,k|q^{\prime },k^{\prime })$ correlation functions are independent
of the length of the surface $L_1$, because they contain $[2\pi\delta (0)]^2$.
At the same time the remaining term in Eq. (4.14), that yields the sum
$C^{(1.5)}(q,k|q^{\prime },k^{\prime })+C^{(2)}(q,k|q^{\prime },k^{\prime
})+C^{(3)}(q,k|q^{\prime }|k^{\prime })$, is inversely proportional to the
surface length, due to the lack of the second delta function. Therefore, in
the limit of a  long surface or a large illumination area the long-range and
infinite-range correlations vanish.

The explicit expressions for the short--range contributions to the angular
intensity correlation functions are
\begin{eqnarray}
C^{(1)}(q,k|q^{\prime },k^{\prime }) = \epsilon_0\frac{16\e_0}{L_1^2}\left( \frac \omega c%
\right) ^2\alpha _0(q)|G(q)|^2\alpha _0(k)|G(k)|^2|\langle
t(q|k)t^{*}(q^{\prime }|k^{\prime })\rangle |^2
\alpha _0(q^{\prime })|G(q^{\prime })|^2\alpha _0(k^{\prime
})|G(k^{\prime })|^2
\lb{c1}
\end{eqnarray}
\begin{eqnarray}
C^{(10)}(q,k|q^{\prime },k^{\prime }) = \epsilon_0\frac{16\e_0}{L_1^2}\left( \frac \omega
c\right) ^2\alpha _0(q)|G(q)|^2\alpha _0(k)|G(k)|^2|\langle
t(q|k)t(q^{\prime }|k^{\prime })\rangle |^2
\alpha _0(q^{\prime })|G(q^{\prime })|^2\alpha _0(k^{\prime
})|G(k^{\prime })|^2,
\lb{c10}
\end{eqnarray}
while the normalized angular intensity correlation functions are given by
\begin{equation}
\Xi ^{(1)}(q,k|q^{\prime },k^{\prime })=\frac{\langle |\langle t(q|k)t^{*}(q^{\prime
}|k^{\prime })\rangle |^2}{\langle |t(q|k)|^2\rangle \langle |t(q^{\prime
}|k^{\prime })|^2\rangle },
\lb{xi1}
\end{equation}
\begin{equation}
\Xi ^{(10)}(q,k|q^{\prime },k^{\prime })=\frac{|\langle t(q|k)t(q^{\prime
}|k^{\prime })\rangle |^2}{\langle |t(q|k)|^2\rangle \langle |t(q^{\prime
}|k^{\prime })|^2\rangle }.
\lb{xi10}
\end{equation}

The property of a speckle pattern that is characterized by the presence of
the factor $2\pi \delta (q-k-q^{\prime }+k^{\prime })$ in $%
C^{(1)}(q,k|q^{\prime },k^{\prime })$ is that if we change the angle of
incidence in such a way that $k$ goes into $k^{\prime }=k+\Delta k$, the
entire speckle pattern shifts in such a way that any feature initially at $q$
moves to $q^{\prime }=q+\Delta k$. This is the reason why the $C^{(1)}$
correlation function was originally named the memory effect.  In terms of the
angles of incidence and scattering, we have that if $\theta _0$ is changed
to $\theta _0^{\prime }=\theta _0+\Delta \theta _0$, any feature in the
speckle pattern originally at $\theta _s$ is shifted to $\theta _s^{\prime
}=\theta _s+\Delta \theta _s$, where $\Delta \theta _s=\Delta \theta _0(\cos
\theta _0/\cos \theta _s)$ to first order in $\Delta \theta _0$.

The property of a speckle pattern that is characterized by the presence of
the factor $2\pi \delta (q-k+q^{\prime }-k^{\prime })$ in $%
C^{(10)}(q,k|q^{\prime },k^{\prime })$ is that if we change the angle of
incidence in such a way that $k$ goes into $k^{\prime }=k+\Delta k$, a
feature at $q=k-\Delta q$ will be shifted to $q^{\prime }=k^{\prime }+\Delta
q$, i.e. to a point as much to one side of the new specular direction as the
original point was on the other side of the original specular direction. For
one and the same incident beam the  $C^{(10)}$ correlation function therefore reflects
the ''symmetry'' of the speckle pattern with respect to the specular
direction.

In this paper we will be concerned only with the short-range correlation
functions. The extraction of $C^{(1.5)},C^{(2)}$, and $C^{(3)}$ from the
third term on the right hand side of \eq{dt} will be described in a
separate paper.

\setcounter{section}{5}
\setcounter{equation}{0}
\section*{5. The Correlation functions $C^{(1)}$ and  $\Xi ^{(1)}$.}

It has been shown in Ref. [14] that if we rewrite the
product $t(q|k)t^{*}(q^{\prime }|k^{\prime })$ in a direct product notation
as $\tau ^{(1)}(q,q^{\prime }|k,k^{\prime })$ the ensemble average of $\tau
^{(1)}(q,q^{\prime }|k,k^{\prime })$ satisfies the equation
\begin{eqnarray}
\langle\tau ^{(1)}(q,q^{\prime }|k,k^{\prime })\rangle  =\langle \Gamma
^{(1)}(q,q^{\prime }|k,k^{\prime })\rangle +\int_{-\infty }^\infty \frac{dp}{%
2\pi }\int_{-\infty }^\infty \frac{dp^{\prime }}{2\pi }\langle \Gamma
^{(1)}(q,q^{\prime }|p,p^{\prime })\rangle
G(p)G^{*}(p^{\prime })\langle \tau ^{(1)}(p,p^{\prime }|k,k^{\prime
})\rangle .
\lb{etau1}
\end{eqnarray}
In this equation $\langle \Gamma ^{(1)}(q,q^{\prime }k,k^{\prime })\rangle $
is an irreducible vertex function. In the present work we will approximate
it by the sum of all  maximally-crossed diagrams  in the small
roughness approximation. It is the contributions
associated with these diagrams that describe the phase-coherent
multiple-scattering processes that give rise to the effects we seek.
Due to the stationarity
of the surface profile function $\zeta (x_1)$ each
term in the expansion of $\langle \Gamma ^{(1)}(q,q^{\prime
}|k,k^{\prime })\rangle $ is proportional to $2\pi \delta (q-k-q^{\prime
}+k^{\prime })$, so that $\langle \Gamma ^{(1)}(q,q^{\prime
}|k,k^{\prime })\rangle $ is given by
\begin{eqnarray}
\langle \Gamma ^{(1)}(q,q^{\prime }|k,k^{\prime })\rangle &=& 2\pi \delta
(q-k-q^{\prime }+k^{\prime })\{W^2g(|k-q|)  \nonumber \\
&&\quad + \int_{-\infty }^\infty \frac{dp_1}{2\pi }
W^2g(|q-p_1|)G(p_1)G^{*}(q^{\prime }+k-p_1)W^2g(|p_1-k|)  \nonumber \\
&&\quad +\int_{-\infty }^\infty \frac{dp_1}{2\pi }\int_{-\infty
}^\infty \frac{dp_2}{2\pi }W^2g(|q-p_2|)G(p_2)G^{*}(q^{\prime
}+k-p_2)W^2g(|p_2-p_1|)G(p_1)   \nonumber \\
&&\qquad \qquad \times G^{*}(q^{\prime }+k-p_1)W^2g(|p_1-k|)+\cdots \}.
\lb{gamma1}
\end{eqnarray}
To evaluate this sum we proceed as follows. We write the power spectrum of
the surface roughness $g(|q-k|)$ in the separable form
\sctr{1}
\begin{eqnarray}
g(|q-k|) =\sqrt{\pi }a\exp\left[-\frac{a^2(q-k)^2}{4}\right]
=\sum_{\ell =0}^\infty \phi _\ell (q)\phi _\ell (k),
\lb{fact}
\end{eqnarray}
where
\sctr{2}\sceq
\begin{eqnarray}
\phi _\ell (q)=\left( \frac{\sqrt{\pi }a^{2\ell +1}}{2^\ell \ell !}\right) ^{%
\frac 12}q^\ell e^{-\frac{a^2}4q^2}.
\end{eqnarray}
In numerical calculations we will replace the upper limit on the sum in
\eq{fact} by an integer $N$, which is increased until a convergent result for $%
\langle \Gamma ^{(1)}(q,q^{\prime }|k,k^{\prime })\rangle $ is obtained.
With the use of this representation we find that
\sctr{0}
\begin{eqnarray}
\langle \Gamma ^{(1)}(q,q^{\prime }|k,k^{\prime })\rangle &=& 2\pi \delta
(q-k-q^{\prime }+k^{\prime })\nn\\
&&\qquad \times\left\{W^2g(|q-k|) 
+ W^2\sum_{\ell =0}^N\sum_{\ell ^{\prime }=0}^N\phi _\ell (q)\{[{\bf I}-%
{\bf K}(q^{\prime }+k)]^{-1}{\bf K}(q^{\prime }+k)]\}_{\ell \ell ^{\prime
}}\phi _{\ell ^{\prime }}(k)\right\} ,
\lb{gamma11}
\end{eqnarray}
where the elements of the $(N+1)\times (N+1)$ matrix ${\bf K}(Q)$ are given
by
\begin{eqnarray}
K_{\ell \ell ^{\prime }}(Q)=W^2\int_{-\infty }^\infty \frac{dp}{2\pi }\phi
_\ell (p)G(p)G^{*}(Q-p)\phi _{\ell ^{\prime }}(p).
\lb{k}
\end{eqnarray}

The result given by \eq{gamma11} now has to be substituted into \eq{etau1},
which is then solved by iteration. However, in each of the integral terms in
the iterative solution we keep only the contribution associated with the
first term on the right hand side of \eq{gamma11}, and omit all contributions
that contain the second term.
The sum of the resulting integral terms is given by
\begin{eqnarray}
&& 2\pi\delta (q-k-q^{\prime}+k^{\prime})\left\{ \int^{\infty}_{-\infty}
\frac{dp_1}{2\pi}W^2g(|q-p_1|)G(p_1)G^*(q-q^{%
\prime}-p_1)W^2g(|p_1-k|) \right.  \nonumber \\
&&\hspace{1.35in}  +\int^{\infty}_{-\infty} \frac{dp_1}{2\pi} \int^{\infty}_{-\infty} \frac{dp_2}{2\pi}\;
W^2g(|q-p_2|)G(p_2)G^*(q-q^{\prime}-p_2)W^2g(|p_2-p_1|)G(p_1)
\nonumber \\
&& \hspace*{2.4in}\left. \vphantom{\wc} \times
G^*(q-q^{\prime}-p_1)W^2g(|p_1-k|)+\ldots \right\}.
\end{eqnarray}
This is nothing more than the sum of the contributions associated with all
the ladder diagrams, starting with the two-rung ladder
diagram. We sum this infinite series with the use of the representation
\ef{fact}, and obtain
\begin{eqnarray}
2\pi\delta (q-k-q^{\prime}+k^{\prime})\;W^2\sum^{N}_{\ell = 0}\sum^{N}_{\ell
^{\prime}=0} \phi_{\ell}(q)\{[ {\bf I} - {\bf K} (q-q^{\prime})]^{-1}{\bf K}%
(q-q^{\prime})\}_{\ell\ell ^{\prime}}\phi_{\ell ^{\prime}}(k) .
\end{eqnarray}
This contribution equals that of the second term on the right hand side of
\eq{gamma11} when $q^{\prime}= - k^{\prime}$. Therefore we cannot neglect it
in comparison with the latter contribution. On combining this result with the
one given by \eq{gamma11} we obtain finally our approximation to $\langle
\tau^{(1)}(q,q^{\prime}|k,k^{\prime})\rangle $:
\begin{eqnarray}
\langle \tau^{(1)}(q,q^{\prime}|k,k^{\prime})\rangle &\equiv& \langle
t(q|k)t^*(q^{\prime}|k^{\prime}) \rangle  \nonumber \\
&=& 2\pi \delta (q-k-q^{\prime}+k^{\prime}) \{ W^2 g(|q-k|)   
+ W^2 \sum^{N}_{\ell = 0} \sum^{N}_{\ell
^{\prime}=0}\phi_{\ell}(q) \{ {\bf I} - {\bf K} (q^{\prime}+k)]^{-1}{\bf K}%
(q^{\prime}+k)\}_{\ell\ell ^{\prime}}\phi_{\ell ^{\prime}}(k) \nn\\&&
\hspace{1.25in} 
+W^2 \sum^{N}_{\ell = 0} \sum^{N}_{\ell
^{\prime}=0}\phi_{\ell}(q) \{ {\bf I} - {\bf K} (q-q^{\prime})]^{-1} {\bf K}%
(q-q^{\prime})\}_{\ell\ell ^{\prime}}\phi_{\ell ^{\prime}}(k)\} .
\lb{tau1}
\end{eqnarray}
The substitution of this result into \eq{c1} yields our result for $%
C^{(1)}(q,k|q^{\prime},k^{\prime})$.
To normalize the correlation function we use the fact that
$\la t(q|k)t^*(q|k)\ra =\tau^{(1)}(q,q|k,k)$.

\setcounter{section}{6}
\setcounter{equation}{0}
\section*{6. The Correlation Functions $C^{(10)}$ and $\Xi^{(10)}$}

In exactly the same way as \eq{etau1} was derived in Ref.\ 14 the correlation function
 $\langle\tau^{(10)}(q,q^{\prime}|k,k^{\prime})\rangle$ $=
\langle \tau (q|k)t(q^{\prime}|k^{\prime})\rangle$ can be shown to be the
solution of the equation
\sctr{0}
\begin{eqnarray}
\langle \tau^{(10)}(q,q^{\prime}|k,k^{\prime}) \rangle = \langle
\Gamma^{(10)}(q,q^{\prime}|k,k^{\prime}) \rangle + \int^{\infty}_{-\infty} \frac{dp}{%
2\pi} \int^{\infty}_{-\infty} \frac{dp^{\prime}}{2\pi} \Gamma^{(10)}(p,p^{%
\prime}|p,p^{\prime}) \rangle
 G(p)G(p^{\prime})\langle
\tau^{(10)}(q,q^{\prime}|k,k^{\prime})\rangle .
\lb{etau10}
\end{eqnarray}
In the present case the averaged irreducible vertex function $%
\langle\Gamma^{(10)}(q,q^{\prime}|k,k^{\prime})\rangle$ will be approximated
in the small roughness limit by the sum of the maximally crossed diagrams.
Evaluating the contribution associated with the maximally crossed diagrams in
a standard manner, we obtain
for $\langle\Gamma^{(10)}(q,q^{\prime}|k,k^{\prime})\rangle$ the result
\begin{eqnarray}
\langle \Gamma^{(10)}(q,q^{\prime}|k,k^{\prime})\rangle &=& 2\pi\delta
(q-k+q^{\prime}-k^{\prime})\{ W^2g(|q-k|)  \nonumber \\
&& \qquad + \int^{\infty}_{-\infty} \frac{dp_1}{2\pi} W^2
g(|q-p_1|)G(p_1)G(k-q^{\prime}-p_1)W^2(g(|p_1-k|)  \nonumber \\
&&\qquad + \int^{\infty}_{-\infty} \frac{dp_1}{2\pi} \int^{\infty}_{-\infty} \frac{%
dp_2}{2\pi}\;W^2g(|q-p_2|) G(k-q^{\prime}-p_2)W^2g(|p_2-p_1|)   \nonumber
\\
&& \hspace{1.5in}\times G(p_1) G(k-q^{\prime}-p_1)W^2g(|p_1-k|)+\cdots \}
\lb{gamma10}
\end{eqnarray}
In writing this expansion we have used the fact that $G(k)$ is an even
function of $k$. We sum this series with the aid of the decomposition
\ef{fact} and obtain
\begin{eqnarray}
\langle \Gamma^{(10)}(q,q^{\prime}|k,k^{\prime}) \rangle 
&=& 2\pi\delta(q-k+q^{\prime}-k^{\prime})\nn\\
&& \qquad \times \left\{ W^2g(|q-k|) 
+W^2 \sum^{N}_{\ell = 0} \sum^{N}_{\ell^{\prime}= 0} \phi_{\ell}(q)%
\left\{[{\bf I} - {\bf L} (k-q^{\prime})]^{-1}{\bf L}(k-q^{\prime})\right%
\}_{\ell\ell ^{\prime}}\phi_{\ell ^{\prime}}(k)\right\} ,
\lb{gamma101}
\end{eqnarray}
where the elements of the $(N+1)\times (N+1)$ matrix ${\bf L}(Q)$ are given
by
\begin{eqnarray}
L_{\ell\ell ^{\prime}}(Q) = W^2 \int^{\infty}_{-\infty} \frac{dp}{2\pi}
\phi_{\ell}(p)G(p)G(Q-p)\phi_{\ell ^{\prime}}(p) .
\lb{L}
\end{eqnarray}
As in the calculation of $\langle \tau^{(1)}(q,q^{\prime}|k,k^{\prime})
\rangle$, when the result given by \eq{gamma101} is substituted into \eq{etau10},
and the resulting integral equation is solved by iteration, in each integral
term in the resulting expansion only the contribution associated with the
first term on the right hand side of \eq{gamma101} is kept, and all
contributions that contain the second are omitted. The sum of the resulting
integral terms is given by
\begin{eqnarray}
&&2\pi\delta (q-k+q^{\prime}-k^{\prime})\left\{ \int^{\infty}_{-\infty} \frac{%
dp_1}{2\pi} W^2g(|q-p_1|)G(p_1)G(q+q^{\prime}-p_1)W^2g(|p_1-k|) \right.  \nonumber
\\
&& \hspace*{1.35in} + \int^{\infty}_{-\infty} \frac{dp_1}{2\pi}%
\int^{\infty}_{-\infty} \frac{dp_2}{2\pi}\;W^2g(|q-p_2|)G(p_2)G(q+q^{%
\prime}-p_2)W^2g(|p_2-p_1) \nonumber \\
&& \hspace*{2.3in} \left. \vphantom{\wc} \times 
   G(p_1)G(q+q^{\prime}-p_1)W^2g(|p_1-k|)+ \cdots \right\}.
\lb{ld10}
\end{eqnarray}
This is just the sum of the contributions associated with all the ladder
diagrams starting with the two-rung ladder diagram. We
sum the infinite series \ef{ld10} with the aid of the representation
\eq{fact} to
obtain the result that it is given by
\begin{eqnarray}
2\pi\delta (q-k+q^{\prime}-k^{\prime})\;W^2\sum^{N}_{\ell = 0}\sum^{N}_{\ell
^{\prime}= 0} \phi_{\ell} (q)\{ [ {\bf I} - {\bf L}(q+q^{\prime})]^{-1}{\bf L%
}(q+q^{\prime})\}_{\ell\ell ^{\prime}}\phi_{\ell ^{\prime}}(k) .
\end{eqnarray}
Thus, our approximation to $\langle\tau^{10}(q,q^{\prime}|k,k^{\prime})%
\rangle$ is
\begin{eqnarray}
\langle\tau^{(10)}(q,q^{\prime}|k,k^{\prime})\rangle &\equiv& 
\langle t(q|k)t(q^{\prime}|k^{\prime})\rangle \nonumber \\
&=& 2\pi\delta (q-k+q^{\prime}-k^{\prime})\left\{ W^2g(|q-k|)  
+ W^2\sum^{N}_{\ell = 0}\sum^{N}_{\ell ^{\prime}=0}\phi_{\ell}(q)\{ [
{\bf I}-{\bf L}(k-q^{\prime})]^{-1}{\bf L}(k-q^{\prime})\}_{\ell \ell
^{\prime}}\phi_{\ell^{\prime}}(k) \right. \nonumber \\
&& %
\hspace{1.5in} 
\left.
+ W^2\sum^{N}_{\ell = 0}\sum^{N}_{\ell ^{\prime}=0}\phi_{\ell}(q)\{ [
{\bf I}-{\bf L}(q+q^{\prime})]^{-1}{\bf L}(q+q^{\prime})\}_{\ell \ell
^{\prime}}\phi_{\ell ^{\prime}}(k) \right\}.
\lb{tau10}
\end{eqnarray}
The second and third terms on the right hand side of this equation are equal
when $k^{\prime}= -k, q^{\prime}=-q,$ and $q = - k$. The substitution of this expression for $%
\langle\tau^{(10)}(q,q^{\prime}|k,k^{\prime})\rangle$ into \eq{c10} yields
our approximation to $C^{(10)}(q,k|q^{\prime},k^{\prime})$.
The correlation function
$\Xi^{(10)}(q,k|q^{\prime},k^{\prime})$  is then readily obtained.

\setcounter{section}{7}
\setcounter{equation}{0}
\section*{7. Results}
To illustrate the results presented in the preceding Sections it is
convenient to represent the correlation functions in the form \sctr{1}
\bqe
C^{(1)}(q,k|q',k')=2\pi\delta(q-k-q'+k')\frac{1}{L_1}C^{(1)}_0(q,k|q',q'-q+k),
\eqe \sctr{2}\sceq \bqe
\Xi^{(1)}(q,k|q',k')=2\pi\delta(q-k-q'+k')\frac{1}{L_1}\Xi^{(1)}_0(q,k|q',q'-q+k),
\eqe and \sctr{1} \bqe
C^{(10)}(q,k|q',k')=2\pi\delta(q-k+q'-k')\frac{1}{L_1}C^{(10)}_0(q,k|q',q'+q-k),
\eqe \sctr{2}\sceq \bqe
\Xi^{(10)}(q,k|q',k')=2\pi\delta(q-k+q'-k')\frac{1}{L_1}\Xi^{(10)}_0(q,k|q',q'+q-k),
\eqe where $C^{(1)}_0(q,k|q',q'-q+k)$, $\Xi^{(1)}_0(q,k|q',q'-q+k)$,
$C^{(10)}_0(q,k|q',q'+q-k)$, and $\Xi^{(10)}_0(q,k|q',q'+q-k)$ are the
envelopes of the correlation functions $C^{(1)}$, $\Xi^{(1)}$,
$C^{(10)}$, and $\Xi^{(10)} $, respectively, and are independent of
the length of the rough surface.  The envelopes are functions of
$\theta_s^{\prime}$ for fixed values of $\theta_0$ and $\theta_s$,
while $\theta_0^{\prime}$ is determined by the constraint of the
$\delta-$function entering the expression for the respective
correlation functions.  In Fig. 2(a) we plot
$C^{(1)}_0(q,k|q',q'-q+k)$ (solid line) and
$C^{(10)}_0(q,k|q',q'+q-k)$ (dashed line) when $s-$polarized light of
wavelength $\lambda= 632.8nm$ is incident on a randomly rough
dielectric surface from vacuum ($\epsilon_0=1$).  The roughness
parameters are $\delta=20 nm$ and $a=100 nm$.  The dielectric constant
of the scattering medium is $\epsilon =2.69$.  For the parameters of
the scattering system assumed the envelopes of both $C^{(1)}$ and
$C^{(10)}$ are structureless functions of $\theta_s^{\prime}$ (the
angles $\theta_s=8^o$ and $\theta_0=3^o$ are fixed, while the angle
$\theta_0^{\prime}$ is determined by the constraint of the
$\delta$-function), and are of almost the same magnitude. The reason
for this is that the scattering from a weakly rough dielectric surface
is extremely weak, and no coherent effects can be observed for such a
system.  The results for the envelopes
of the normalized\\
\vskip-2.in
\hbox{\psfig{figure=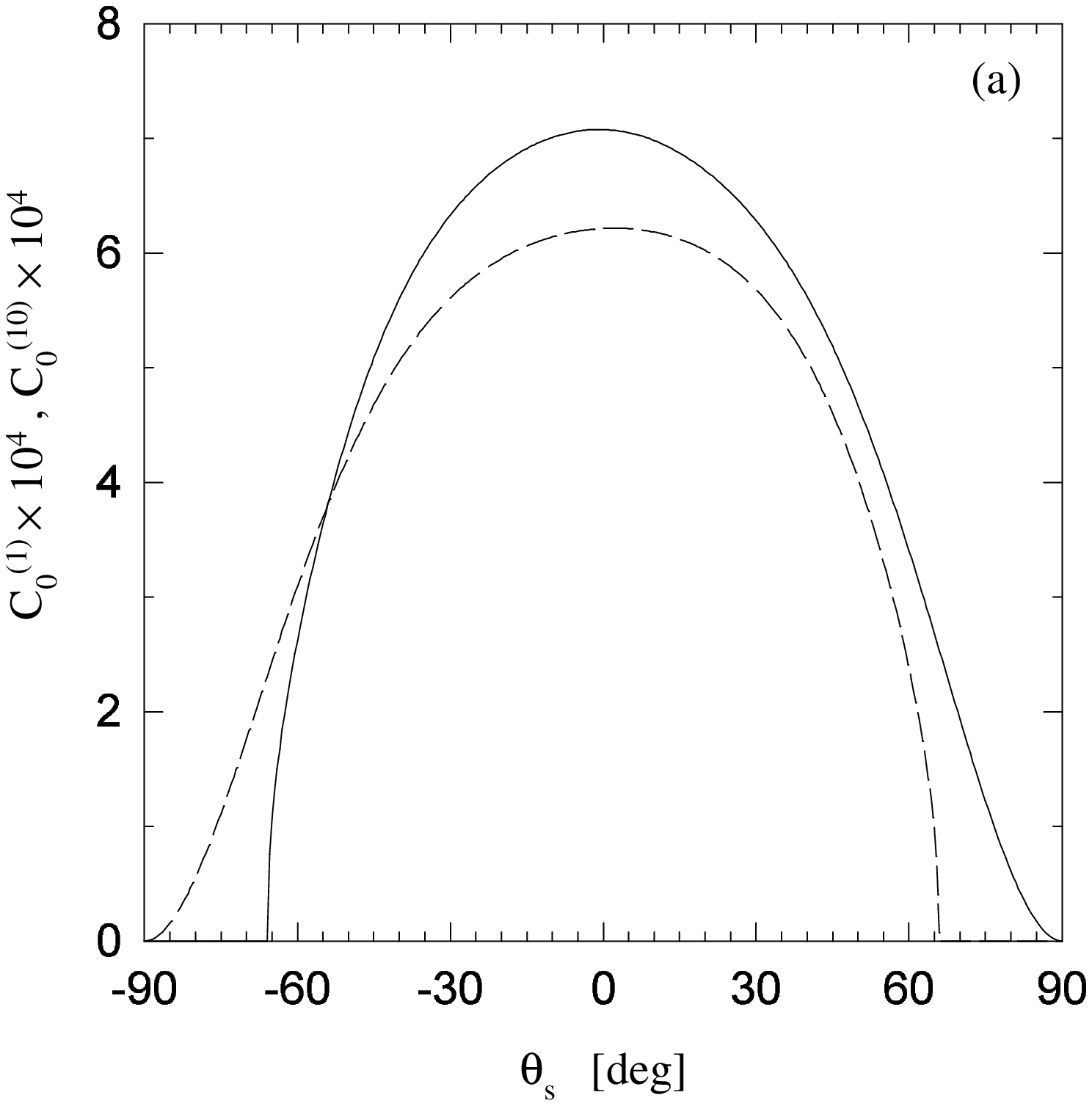,height=3in}\hskip.5in\psfig{figure=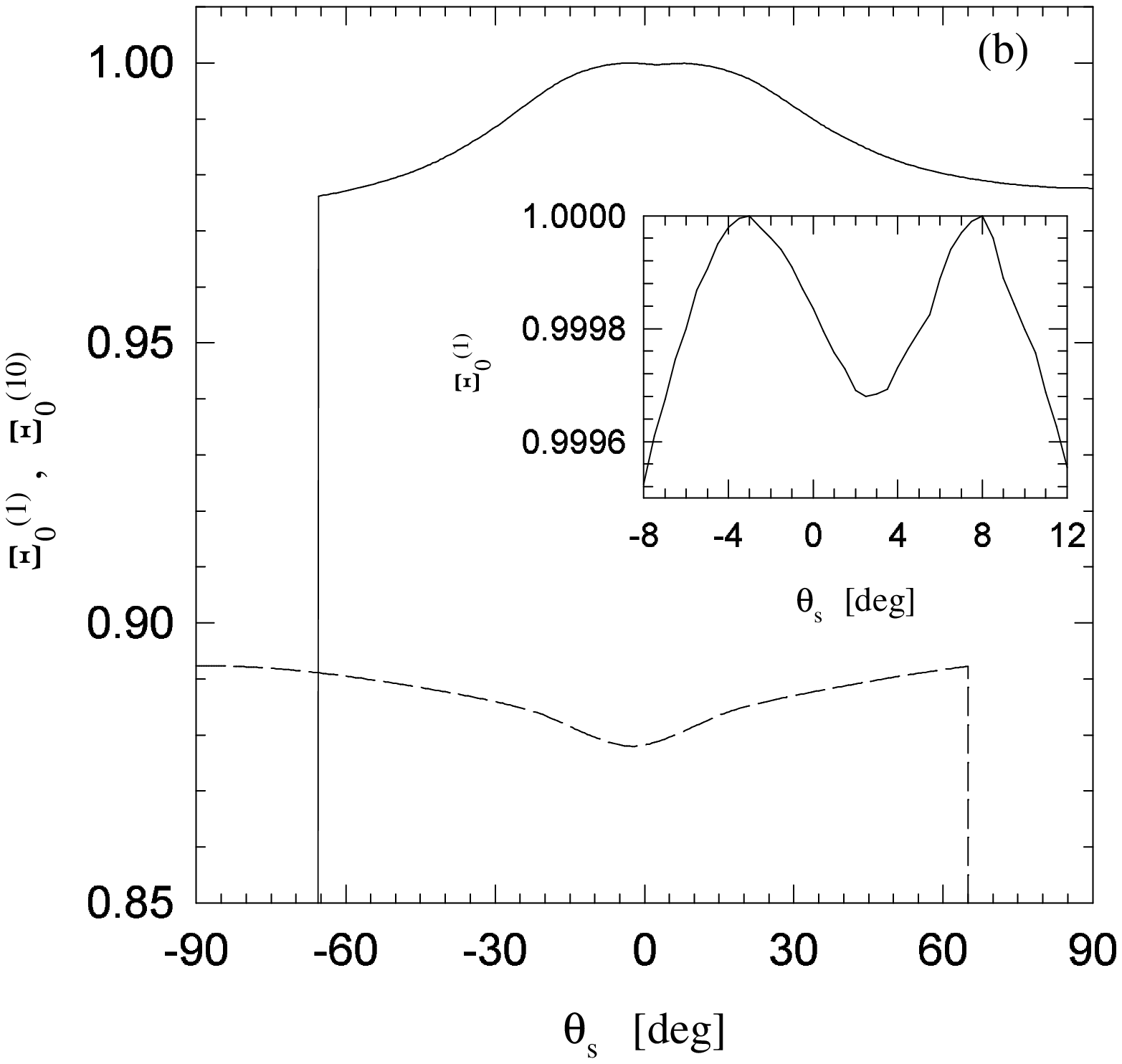,height=3in}}
\vskip-2.5in
\begin{itemize}
\item[Fig. 2.]The envelopes of the $C^{(1)}$ (solid line) and $C^{(10)}$
(dashed line) correlation functions (a) and $\Xi^{(1)}$ (solid line) and $\Xi^{(10)}$
(dashed line) correlation functions (b) as functions of $\theta_s^{\prime}$
when $s-$polarized light is scattered from a randomly rough surface of
a dielectric medium with $\e = 2.69$.
\end{itemize}
\noindent
correlation functions $\Xi^{(1)}$ (solid line) and $\Xi^{(10)}$ (
dashed line) obtained for the same parameters are plotted in Fig.
2(b). The function $\Xi^{(1)}$ shows perfect correlations in the
vicinity of the memory and reciprocal memory effect peaks, although
the envelope of $C^{(1)}$ does not display any peaks.  The minimum in
the plot of $\Xi^{(10)}_0(q,k|q',q'+q-k)$ occurs in the vicinity of
the backscattering direction and is due to its normalization.

As is known [16], when light is scattered by a rough interface between
two dielectric media with low dielectric constrast, although the
scattering is weak, nevertheless, an enhanced backscattering peak can
be formed in the angular dependence of the intensity of the light
scattered incoherently, due to the coherent interference of the direct
and reciprocal paths traversed by multiply scattered lateral waves.
When such an enhanced backscattering peak occurs one can expect peaks
associated with the memory and reciprocal meanory effects to appear in
$C^{(1)}_0(q,k|q^{\prime},q^{\prime}-q +k)$.  This is indeed what is
found. In Fig. 3(a) the envelopes of the correlation functions
$C^{(1)}$ (solid line) and $C^{(10)}$ (dashed line) are plotted as
functions of $\theta_s^{\prime}$ when $s-$polarized light of
(vacuum) wavelength $\lambda=632.8nm$ is incident
on the rough interface between two dielectrics characterized by the
dielectric constants $\e_0=2.6$ and $\e=2.69$.  The roughness
parameters are the same ones assumed in obtaining the results plotted
in Fig. 2.  The envelope of $C^{(1)}$ displays peaks at $q'=q$ (memory
effect) and at $q'=-k$ (reciprocal memory effect), while the envelope
of $C^{(10)}$ remains structureless, but smaller in its amplitude.
The envelopes of the normalized correlation functions $\Xi^{(1)}$
(Fig. 3(b), solid line) and $\Xi^{(10)}$ (Fig. 3(b), dashed line) show
that perfect correlations occur in the vicinity of the memory and
reciprocal memory effect peaks.  \newpage
\hbox{\psfig{figure=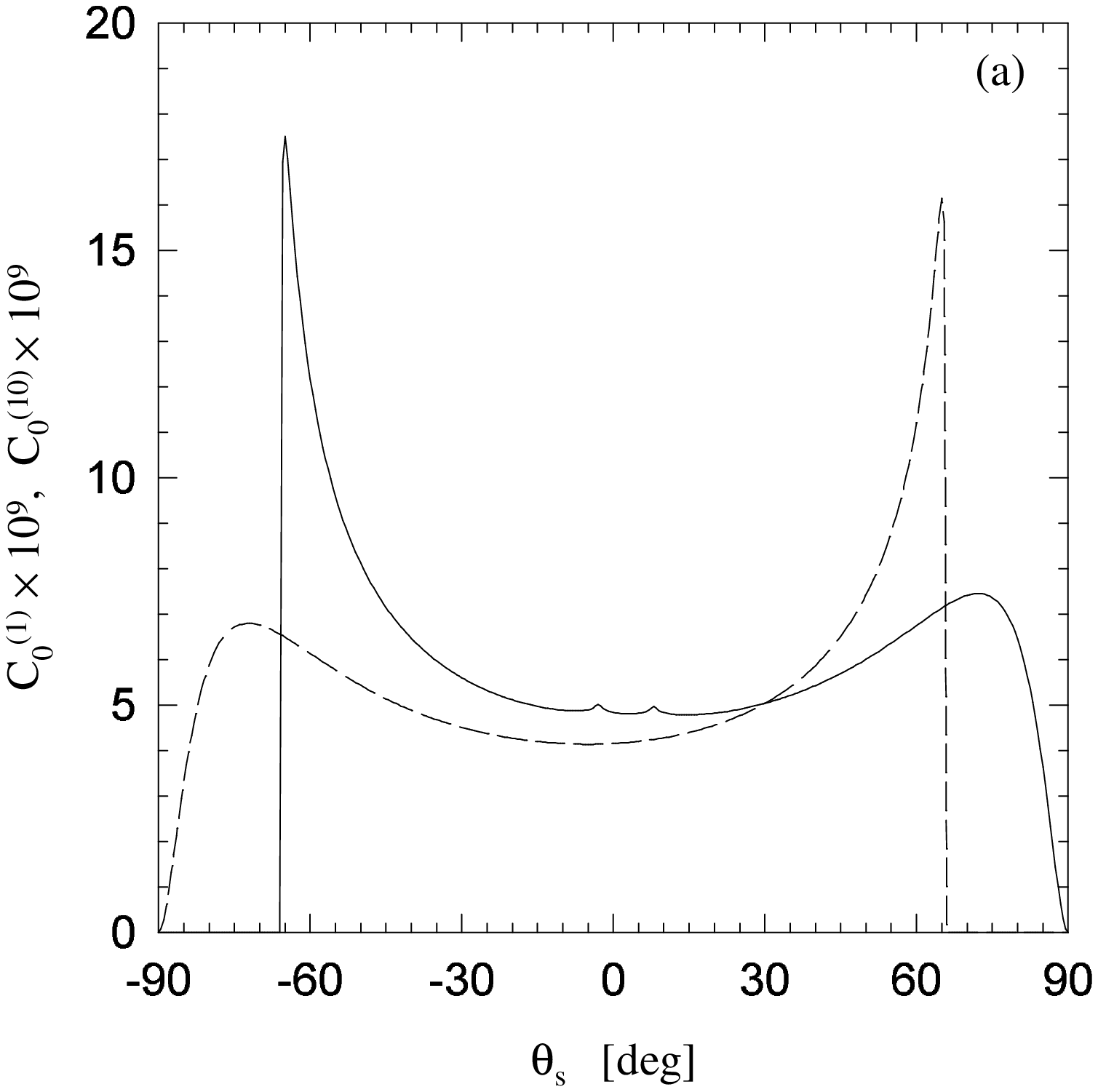,height=3in}\hskip.5in\psfig{figure=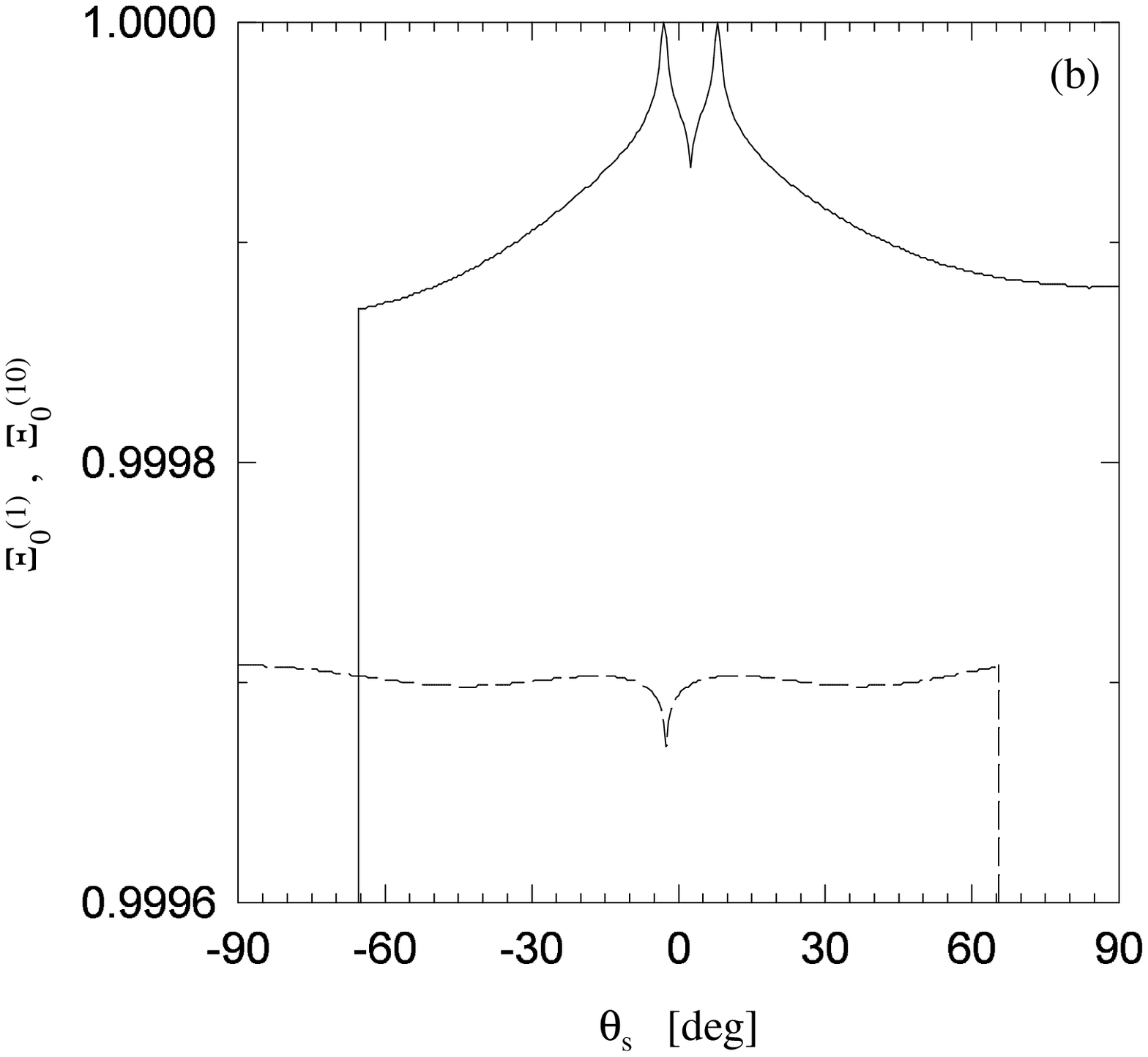,height=3in}}
\vskip-2.5in
\begin{itemize}
\item[Fig. 3.]
The envelopes of the $C^{(1)}$ (solid line) and $C^{(10)}$
(dashed line) correlation functions (a) and  $\Xi^{(1)}$ (solid line) and $\Xi^{(10)}$
(dashed line) correlation functions (b) as functions of $\theta_s^{\prime}$
when $s-$polarized light is scattered from a randomly rough interface between
two dielectric media with $\e_0 = 2.6$  and $\e=2.69$.
\end{itemize}

\setcounter{section}{8}
\setcounter{equation}{0}
\section*{8. Conclusions}

In this paper we have presented a theoretical study of the angular intensity
correlation functions $C^{(1)}$ and $C^{(10)}$ of s-polarized light
scattered from a one-dimensional random interface between two semi-infinite
dielectric media. The angular intensity correlation functions
are calculated by means of an approach that
explicitly separates out the $C^{(1)}$ and $C^{(10)}$ contributions to it.
We have shown that in the case of a random vacuum-dielectric interface
the envelopes of  $C^{(1)}$ and $C^{(10)}$ are structureless
functions of the scattering angle $\theta_s^{\prime}$ of almost the
same magnitude, nevertheless, the envelope of the normalized
correlation function $\Xi^{(1)}$ shows perfect correlations at the
positions of the peaks of the memory and reciprocal memory effects.
The envelope of the normalized correlation function $\Xi^{(10)}$
displays a minimum at the position of the enhanced backscattering peak, which is
due to its normalization.

It is also shown that in the case of a random interface between two
dielectrics with a low dielectric contrast
the envelope of $C^{(1)}$ displays the peaks of the memory and reciprocal
memory effects.

\acknowledgements  The work of T .A. L. and A. A. M. was
supported in part by Army Research Office Grant No. DAAG 55-98-C-0034.
I. S. would like to thank the Research Council of Norway
(Contract No. 32690/213) and Norsk Hydro ASA for financial support.
\newpage
\begin{center}
{\bf  REFERENCES}
\end{center}
\begin{enumerate}
\item V. Malyshkin, A.R. McGurn, T.A. Leskova, A.A. Maradudin, and
M. Nieto--Vesperinas, Opt. Lett. {\bf 22}, 946 (1997).
\item V. Malyshkin, A.R. McGurn, T.A. Leskova, A.A. Maradudin, and
M. Nieto--Vesperinas, Waves in Random Media {\bf 7}, 479 (1997).
\item  S. Feng, C. Kane, P.A. Lee, and A.D. Stone, Phys. Rev. Lett. {\bf
61}, 834 (1988).
\item I.\ Freund, M.\ Rosenbluh, and S.\ Feng, 
         Phys. Rev. Lett. {\bf 61}, 2328 (1988);
      I.\ Freund, M.\ Rosenbluh, and R.\ Berkovits,
         Phys. Rev. B{\bf 39}, 12403 (1989);
      I.\ Freund and R.\ Berkovits,
         Phys. Rev. B{\bf 41}, 496 (1990) (erratum  {\bf 41}, 9540
         (1990));
      I.\ Freund and M.\ Rosenbluh, Opt. Commun. {\bf 82}, 362 (1991).
\item N.\ Garcia and A.\ Z.\ Genack, 
         Phys. Rev. Lett. {\bf 63}, 1678 (1989);
      M.\ P.\ van Albada, J.\ F.\ de Boer, and A.\ Lagendijk,
         Phys. Rev. Lett. {\bf 64}, 2787 (1990).
\item F.\ Scheffold and G.\ Maret, 
         Phys.\ Rev.\ Lett.\ {\bf 81}, 5800 (1998).
\item A.\ Arsenieva and S.\ Feng, Phys.\ Rev.\ B{\bf 47}, 13047 (1993).
\item B. Shapiro, Phys. Rev. Lett. {\bf 57}, 2168 (1986).
\item C.\ S.\ West and K.\ A.\ O'Donnell, Phys. Rev. B{\bf 59}, 2393 (1999).
\item Jun Q. Lu and Zu-Han Gu, Appl. Opt. {\bf 36}, 4562 (1997); Zu-Han Gu and Jun Q. Lu, SPIE {\bf 3141}, 269 (1997).
\item T. Tamir, in: {\it Electromagnetic Surface Modes}, ed. A. D. Boardman
(J.Wiley \& Sons, New York, 1992), p. 521.
\item A.\ A.\ Maradudin, R.\ E.\ Luna, and E.\ R.\ M\'endez, 
         Waves in Random Media {\bf 3}, 51 (1993).
\item F.\ Toigo, A.\ Marvin, V.\ Celli, and N.\ R.\ Hill, 
         Phys. Rev. B{\bf 15}, 5618 (1977).
\item G. C. Brown, V. Celli, M. Haller, A.A. Maradudin, and A. Marvin,
Phys. Rev. B{\bf 31}, 4993 (1985).
\item R. Kubo, J. Phys. Soc. Japan {\bf 17}, 1100 (1962).
\item T. Kawanishi, H. Ogura, and Z.\ L.\ Wang, Waves in Random Media
{\bf 7}, 351 (1997).
\end{enumerate}
\end{document}